\documentclass[twocolumn,showpacs,aps,prl,superscriptaddress,letterpaper]{revtex4}

\usepackage{graphicx}
\usepackage{dcolumn}
\usepackage{amsmath}
\usepackage{epsfig}
\usepackage{multirow}

\long\def\inst#1{\par\nobreak\kern 4pt\nobreak
    {\itshape #1}\par\vskip 10pt plus 3pt minus 3pt}
\RequirePackage{xspace}
\usepackage{relsize}
\def\qqbar {\ensuremath{q\overline q}\xspace}
\def\babar{\mbox{\slshape B\kern-0.1em{\smaller A}\kern-0.1em
    B\kern-0.1em{\smaller A\kern-0.2em R}}}
\def\Abar    {\kern 0.18em\overline{\kern -0.18em A}{}\xspace}
\def\Kbar    {\kern 0.18em\overline{\kern -0.18em K}{}\xspace}
\def\Bbar    {\kern 0.18em\overline{\kern -0.18em B}{}\xspace}

\def\BB      {\ensuremath{B\Bbar}\xspace} 
\def\Bz      {\ensuremath{B^0}\xspace}
\def\Bzb     {\ensuremath{\Bbar^0}\xspace}
\def\BzBzb   {\ensuremath{\Bz {\kern -0.16em \Bzb}}\xspace}
\def\Bu      {\ensuremath{B^+}\xspace}
\def\Bub     {\ensuremath{B^-}\xspace}

\def\BpBm    {\ensuremath{\Bu {\kern -0.16em \Bub}}\xspace}

\newcommand{\optbar}[1]{\shortstack{{\tiny (\rule[.4ex]{1em}{.1mm})}
  \\ [-.7ex] $#1$}}
\def\BorBbar    {\kern 0.18em\optbar{\kern -0.18em B}{}\xspace}
\def\DorDbar    {\kern 0.18em\optbar{\kern -0.18em D}{}\xspace}
\def\KorKbar    {\kern 0.18em\optbar{\kern -0.18em K}{}\xspace}

\def\pep2{PEP-II}
\mathchardef\Upsilon="7107
\def\Y#1S{\ensuremath{\Upsilon{(#1S)}}\xspace}

\def\FourS {\Y4S}

\newcommand{\BABARPubYear}     {04}
\newcommand{\BABARPubNumber}  {031}

\newcommand{\SLACPubNumber} {10564}

\begin{document}

\begin{flushleft}
\babar-PUB-\BABARPubYear/\BABARPubNumber\\
SLAC-PUB-\SLACPubNumber
\\[10mm]
\end{flushleft}

\title{
\large \bfseries \boldmath
Measurement of the $B^0\to\phi K^{*0}$ Decay Amplitudes
}

%
\author{B.~Aubert}
\author{R.~Barate}
\author{D.~Boutigny}
\author{F.~Couderc}
\author{J.-M.~Gaillard}
\author{A.~Hicheur}
\author{Y.~Karyotakis}
\author{J.~P.~Lees}
\author{V.~Tisserand}
\author{A.~Zghiche}
\affiliation{Laboratoire de Physique des Particules, F-74941 Annecy-le-Vieux, France }
% ...
\collaboration{The \babar\ Collaboration}
\noaffiliation

\date{August 6, 2004}


\begin{abstract}
With a sample of about 227 million $\BB$ pairs recorded 
with the $\babar$ detector at the PEP-II storage ring we 
perform a full angular analysis of the decay
$B^0\to\phi K^{*0}(892)$.
We measure the branching fraction to be
$(9.2\pm{0.9}\pm 0.5)\times 10^{-6}$
and determine the fractions of longitudinal and parity-odd 
transverse contributions as
${f_L}=0.52\pm{0.05}\pm 0.02$ and 
${f_\perp}=0.22\pm{0.05}\pm 0.02$, respectively.
The phases of the parity-even and parity-odd transverse 
amplitudes relative 
to the longitudinal amplitude are found to be
${\phi_\parallel}=2.34^{+0.23}_{-0.20}\pm 0.05$ rad and
${\phi_\perp}=2.47\pm 0.25\pm 0.05$ rad, respectively.
We measure five $C\!P$ asymmetries which provide 
important limits on $C\!P$ violation originating
from new physics. We also observe the decay
$B^0\to\phi K^{*0}(1430)$.
\end{abstract}

\pacs{13.25.Hw, 11.30.Er, 12.15.Hh}

\maketitle


The decay $B\to\phi K^*(892)$ is expected to have contributions
from $b\to s$ loop transitions while the tree-level transition
is suppressed in the Standard Model. Angular correlation 
measurements and asymmetries are particularly sensitive to 
amplitudes arising outside the Standard Model~\cite{bvv1}. 
The first evidence for this decay was provided by 
the CLEO~\cite{cleo:phikst} and \babar~\cite{babar:phikst} 
experiments. The large fraction of transverse polarization
observed by \babar~\cite{babar:vv} and confirmed by 
BELLE~\cite{belle:phikst} 
enables a full angular analysis described by 
ten parameters for contributing amplitudes 
and their relative phases.


The angular distribution of the $B\to\phi K^*$ decay products 
can be expressed as a function of
${\cal H}_i=\cos\theta_i$ and $\Phi$, where $\theta_i$ 
is the angle between the direction
of the $K$ from the $K^*\to K\pi$ ($\theta_1$) 
or $\phi\to K\Kbar$ ($\theta_2$) 
and the direction opposite the $B$ in the 
vector resonance rest frame,
and $\Phi$ is the angle between the two resonance decay planes.
The differential decay width has three 
complex amplitudes $A_\lambda$ corresponding to the vector 
meson helicity $\lambda=0$ or $\pm 1$~\cite{bvv1, bvv2}. 
When the last two are expressed in terms
of $A_{\parallel}=(A_{+1}+A_{-1})/\sqrt{2}$ and
$A_{\perp}=(A_{+1}-A_{-1})/\sqrt{2}$ we have
\begin{eqnarray}
\label{eq:helicityfull}
{{8\pi} \over {9\Gamma}}~\!{d^3\Gamma \over 
d{\cal H}_1 d{\cal H}_2d\Phi} 
= {1\over|A_0|^2 + |A_{\parallel}|^2 + |A_{\perp}|^2} \times 
\LARGE\{~
\nonumber \\
 |A_0|^2 ~\!{\cal H}_1^2~\!{\cal H}_2^2
+{1\over 4}~\!(|A_{\parallel}|^2 + |A_{\perp}|^2)
   ~\! (1-{\cal H}_1^2)~\!(1-{\cal H}_2^2)~~
\nonumber \\
+{1\over 4}~\!(|A_{\parallel}|^2 - |A_{\perp}|^2)
    ~\! (1-{\cal H}_1^2)~\!(1-{\cal H}_2^2)~\!\cos2\Phi~~
\nonumber \\
 -{\rm Im}(A_{\perp}A^*_{\parallel})
    ~\! (1-{\cal H}_1^2)~\!(1-{\cal H}_2^2)~\!\sin2\Phi~~
\nonumber \\
 +{\sqrt{2}}~\!{\rm Re}(A_{\parallel}A^*_0)
~\!{\cal H}_1~\!{\cal H}_2
~\!\sqrt{1-{\cal H}_1^2}~\!\sqrt{1-{\cal H}_2^2}
~\!\cos\Phi~~~
\nonumber \\
-{\sqrt{2}}~\!{\rm Im}(A_{\perp}A^*_0)
~\!{\cal H}_1~\!{\cal H}_2
~\!\sqrt{1-{\cal H}_1^2}~\!\sqrt{1-{\cal H}_2^2}
~\!\sin\Phi{~\LARGE\} }.
\end{eqnarray}

In this analysis, we measure the branching fraction,
obtained from the number of reconstructed signal 
events $n_{\rm sig}$, the polarization parameters 
$f_L={|A_0|^2/\Sigma|A_\lambda|^2}$,
$f_{\perp}={|A_{\perp}|^2/\Sigma|A_\lambda|^2}$,
and the relative phases
$\phi_{\parallel} = {\rm arg}(A_{\parallel}/A_0)$,
$\phi_{\perp} = {\rm arg}(A_{\perp}/A_0)$.
We allow for $C\!P$-violating differences between
the $\Bbar^0$ ($Q=+1$) and ${B}^0$ ($Q=-1$) decay 
amplitudes ($\Abar_{\lambda}$ and $A_{\lambda}$), 
where the flavor sign $Q$ is determined 
in the self-tagging final state with 
a $\Kbar^{*}$~or~$K^{*}$:
\begin{eqnarray}
\label{eq:observe}
n_{\rm sig}^Q=n_{\rm sig}~\!(1+Q~\!{\cal A}_{C\!P})/2;
~~~~~~~~~~~~~~~~\\
f_{L}^{~\!Q} = f_{L}~\!(1+Q~\!{\cal A}_{C\!P}^{0}); ~~~~
f_{\perp}^{~\!Q} = f_{\perp}~\!(1+Q~\!{\cal A}_{C\!P}^{\perp});
\nonumber \\
\phi_{\parallel}^Q = \phi_{\parallel}+Q~\!\Delta \phi_{\parallel}; ~~~~
\phi_{\perp}^Q = \phi_{\perp}+{\pi\over 2}
 + Q~\!(\Delta \phi_{\perp}+{\pi\over 2}).
\nonumber
\end{eqnarray}

If one loop diagram dominates the 
decay amplitude, the three direct ${C\!P}$ asymmetries 
${\cal A}_{C\!P}$, ${\cal A}_{C\!P}^{0}$, and 
${\cal A}_{C\!P}^{\perp}$, and the two weak-phase differences 
$\Delta\phi_{\parallel}$ and $\Delta\phi_{\perp}$ 
are expected to be negligible. 
From the above parameters one can derive vector triple-product 
asymmetries ${\cal A}_T^{\parallel}$ and ${\cal A}_T^{0}$
as discussed in Ref.~\cite{bvv1}: 
\begin{eqnarray}
\label{eq:tripleprod}
{\cal A}_T^{\parallel,0}= {1\over 2}\left(
{ {\rm Im}(A_{\perp}A^{*}_{\parallel,0}) \over \Sigma|A_\lambda|^2 } +
{ {\rm Im}(\Abar_{\perp}\Abar^{*}_{\parallel,0}) \over \Sigma|\Abar_\lambda|^2 }\right).
\end{eqnarray}


We use data collected with the 
\babar\ detector~\cite{babar} at the \pep2 asymmetric-energy 
$e^+e^-$ collider~\cite{pep} operated at the 
center-of-mass (CM) energy of the $\FourS$ resonance
($\sqrt{s}=10.58$~GeV).
These data represent an integrated luminosity 
of about 205~fb$^{-1}$, corresponding to $226.6\pm2.5$ million 
$\BB$ pairs.

Charged-particle momenta are measured in a tracking system 
consisting of a five-layer double-sided silicon vertex tracker  
and a 40-layer central drift chamber, 
both immersed in a 1.5-T solenoidal magnetic field. 
Charged-particle identification is provided by 
measurements of the energy loss 
(${\rm d}E/{\rm d}x$) in the tracking devices and
by a ring-imaging Cherenkov detector.


We fully reconstruct $\BorBbar^0\to\phi\KorKbar^{*0}$ 
candidates from their decay products $\phi\to K^+K^-$ 
and $\KorKbar^{*0}\to K^\pm\pi^\mp$
as discussed in Ref.~\cite{babar:vv}.
Charged track candidates are required to originate 
from a single vertex near the interaction point.
We identify $B$ meson candidates kinematically
using the beam-energy-substituted mass $m_{\rm{ES}} =$ 
$[{ (s/2 + \mathbf{p}_i \cdot \mathbf{p}_B)^2 / E_i^2 - 
\mathbf{p}_B^{\,2} }]^{1/2}$ and the energy difference
$\Delta{E}=(E_iE_B-\mathbf{p}_i$$\cdot$$\mathbf{p}_B-s/2)/\sqrt{s}$,
where $(E_i,\mathbf{p}_i)$ is the initial state four-momentum
obtained from the beam momenta, and $(E_B,\mathbf{p}_B)$
is the four-momentum of the reconstructed $B$ candidate.
The requirements on the $K^*$ and $\phi$ invariant masses are 
$0.75 < m_{K\!\pi} < 1.05$ and $0.99 < m_{K\!\Kbar} < 1.05$ (GeV).
We move the selection window to $1.13 < m_{K\!\pi} < 1.73$ (GeV)
in the study of the higher-mass $K^*$ resonances.

To reject the dominant quark-antiquark continuum background,
we require $|\cos\theta_T| < 0.8$, where $\theta_T$ 
is the angle between the $B$-candidate thrust axis
and that of the rest of the tracks and neutral clusters in
the event, calculated in the CM frame. 
We also construct a Fisher discriminant, ${\cal F}$,
further discriminating between signal and background,
that combines the following variables: the polar angles 
of the $B$-momentum vector and the $B$-candidate thrust 
axis with respect to the beam axis in the CM frame, 
and the two Legendre moments $L_0$ and $L_2$ of the energy 
flow around the $B$-candidate thrust axis~\cite{bigPRD}.

Contamination from other $B$ decays is small 
(about 2\% of the total background) 
according to Monte Carlo (MC) simulation~\cite{geant} 
and is taken into account in the fit described below. 
We remove signal candidates that have decay products 
with invariant mass within 12 MeV of the nominal mass 
values for $D_s^\pm$ or $D^\pm\to{\phi\pi^\pm}$.


We use an unbinned, extended maximum-likelihood fit to
extract simultaneously the signal yield and angular distributions
from a sample of selected events.
There are several event categories $j$: signal, 
continuum~$\qqbar$, combinatoric $\BB$ background, 
$B\to\phi K\pi$ with a non-resonant S-wave $K^\pm\pi^\mp$ 
contribution, and $B\to f_0(980)K^*$ with a broad S-wave 
$K^+K^-$ contribution.
The likelihood for each candidate~$i$ is defined as
${\cal L}_i = \sum_{j,k}n_{j}^k\, 
{\cal P}_{j}^k(\vec{x}_{i};\vec{\alpha};\vec{\beta})$,
where each of the ${\cal P}_{j}^k(\vec{x}_{i};\vec{\alpha};\vec{\beta})$ is 
the probability density function (PDF) for variables
$\vec{x}_{i}=\{m_{\rm{ES}}$, $\Delta E$, ${\cal F}$, 
$m_{K\!\pi}$, $m_{K\!\Kbar}$, ${\cal H}_1$, ${\cal H}_2$, $\Phi$, $Q$\}.
The flavor index $k$
corresponds to the measured value of $Q$, that is
${\cal P}_{j}^k\equiv{\cal P}_{j}\times\delta_{kQ}$.
The $n_{j}^k$ is the number of events with the flavor $k$
in the category~$j$.

The PDF ${\cal P}_{j}^k(\vec{x}_{i};\vec{\alpha};\vec{\beta})$ 
for a given candidate $i$ is the product of the PDFs
for each of the variables and a joint PDF for the  
helicity angles and resonance masses as discussed below.
The signal angular distributions are parameterized with 
the set $\vec{\alpha}=\{f_L$, $f_{\perp}$, $\phi_{\parallel}$, 
$\phi_{\perp}$, ${\cal A}_{C\!P}^0$, ${\cal A}_{C\!P}^{\perp}$, 
$\Delta \phi_{\parallel}$, $\Delta \phi_{\perp}$\}
which are left free to vary in the fit. The other PDF 
parameters $\vec{\beta}$ are extracted from MC simulation 
and data in $m_{\rm{ES}}$ and $\Delta E$ sidebands
and are fixed in the fit.
The MC resolutions are adjusted by 
comparing data and MC in calibration channels with similar 
kinematics and topology,
such as $B^0\to D^-\pi^+$ with $D^-\to K^+\pi^-\pi^-$. 
The PDF parameterization for each event candidate
accounts for the loss of acceptance near ${\cal H}_1=0.8$ 
due to the $D_s^\pm$ and $D^\pm$ rejection requirements. 

We use a three-dimensional description for the helicity part 
of the signal PDF, using the ideal angular distribution from 
Eq.~(\ref{eq:helicityfull})
multiplied by an acceptance function 
${\cal{G}}({\cal H}_1,{\cal H}_2,\Phi)$
parameterized with empirical polynomial functions.
The detector acceptance effects are found to be 
uniform in $\Phi$, and we factor the ${\cal H}_1$ and 
${\cal H}_2$ dependence as
${\cal{G}}\equiv{\cal{G}}_1({\cal H}_1)\times{\cal{G}}_2({\cal H}_2)$.
We use two Gaussian functions 
for the parameterization of the signal PDFs 
for $\Delta E$, $m_{\rm{ES}}$, and ${\cal F}$.
A relativistic $P$-wave Breit-Wigner distribution,
convoluted with a Gaussian resolution function, 
is used for the resonance masses.

Parameterization of the non-resonant $B$-decay 
contributions is identical to that of the signal 
for $m_{\rm ES}$, $\Delta E$, and ${\cal F}$, 
but is different for the angular and invariant mass 
distributions. 
In particular, a broad invariant mass distribution 
accounts for all potential S-wave contributions leaking 
into the mass selection window.
For the combinatorial background, we use polynomials,
except for $m_{\rm{ES}}$ and ${\cal F}$ distributions
which are parameterized by an empirical 
phase-space function and by the two Gaussian functions,
respectively.
Resonance production occurs in the background and this 
is taken into account in the PDF.
The background ${\cal H}_i$ distribution is separated into 
contributions from combinatorial background and from 
real vector mesons.

\begingroup
\begin{table}[btp]
\caption{Summary of the $B^0\to\phi K^{*0}(892)$ fit results.
We show results for the ten primary signal fit parameters
defined in Eq.~(\ref{eq:observe}) and the derived parameters: 
reconstruction efficiency $\epsilon$ which depends on decay 
polarization, branching fraction ${\cal B}$, and
triple-product asymmetries from Eq.~(\ref{eq:tripleprod}).
All results include systematic errors, which are quoted 
following the statistical errors.
For the dominant correlations we give the coefficients 
in the last column.
}
\label{tab:results}
\begin{center}
\begin{ruledtabular}
\setlength{\extrarowheight}{1.5pt}
\begin{tabular}{ccl}
\vspace{-3mm} & & \\
   Fit parameter 
 & Fit result & Correlation
\cr
\vspace{-3mm} & & \\
\hline
\vspace{-3mm} & & \\
  $n_{\rm sig}$ (events) 
 & $201\pm 20\pm 6$ &  
\cr
\vspace{-3mm} & & \\
  ${f_L}$  
 & $0.52\pm{0.05}\pm 0.02$  & \multirow{2}{13mm}{{\Large\}}~$-46\%$}
\cr
\vspace{-3mm} & & \\
  ${f_\perp}$ 
 & $0.22\pm{0.05}\pm 0.02$  & 
\cr
\vspace{-3mm} & & \\
  ${\phi_\parallel}$ (rad) 
 & $2.34^{+0.23}_{-0.20}\pm 0.05$ &  \multirow{2}{13mm}{{\Large\}}~$+70\%$}
\cr
\vspace{-3mm} & & \\
  ${\phi_\perp}$ (rad) 
 & $2.47\pm 0.25\pm 0.05$  &  
\cr
\vspace{-3mm} & & \\
  ${\cal A}_{C\!P}$ 
 & $-0.01\pm{0.09}\pm 0.02$  & 
\cr
\vspace{-3mm} & & \\
  ${\cal A}_{C\!P}^0$ 
 & $-0.06\pm{0.10}\pm 0.01$   &  \multirow{2}{13mm}{{\Large\}}~$-45\%$}
\cr
\vspace{-3mm} & & \\
  ${\cal A}_{C\!P}^{\perp}$ 
 & $-0.10\pm 0.24\pm 0.05$  &  
\cr
\vspace{-3mm} & & \\
  $\Delta \phi_{\parallel}$ (rad) 
 & $0.27^{+0.20}_{-0.23}\pm 0.05$   &  \multirow{2}{13mm}{{\Large\}}~$+70\%$}
\cr
\vspace{-3mm} & & \\
  $\Delta \phi_{\perp}$ (rad) 
 & $0.36\pm 0.25\pm 0.05$   & 
\cr
\vspace{-3mm} & & \\
\hline
\vspace{-3mm} & & \\
  $\epsilon$ (\%) 
 & $9.7\pm 0.5$  &  
\cr
\vspace{-3mm} & & \\
  ${\cal B}$ 
 & $(9.2\pm{0.9}\pm 0.5)\times 10^{-6}$  &  
\cr
\vspace{-3mm} & & \\
  ${\cal A}_T^{\parallel}$ 
 & $-0.02\pm{0.04}\pm 0.01$  &  
\cr
\vspace{-3mm} & & \\
  ${\cal A}_T^{0}$ 
 & $+0.11\pm{0.05}\pm 0.01$  &  
\cr
\vspace{-3mm} & & \\
\end{tabular}
\end{ruledtabular}
\end{center}
\end{table}
\endgroup

\begin{figure}[hbt]
\centerline{
\setlength{\epsfxsize}{1.0\linewidth}\leavevmode\epsfbox{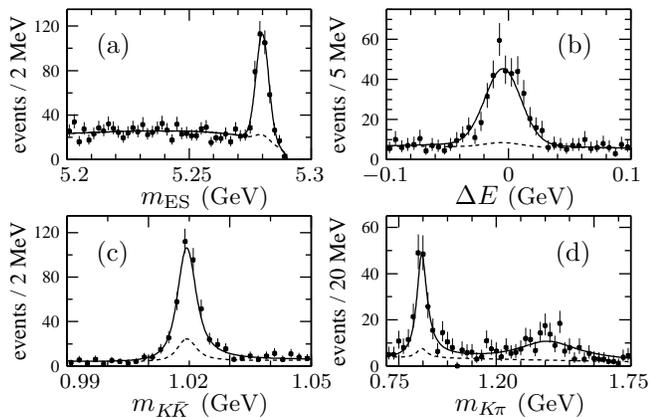}
}
\vspace{-0.3cm}
\caption{\label{fig:projection1} 
Projections onto the variables 
$m_{\rm ES}$ (a), $\Delta E$ (b), $m_{K\!\Kbar}$ (c), and
$m_{K\!\pi}$ (d) for the signal $B^0\to\phi K^{*0}(892)$
and $\phi K^{*0}(1430)$ candidates combined.
}
\end{figure}
\begin{figure}[hbt]
\centerline{
\setlength{\epsfxsize}{1.0\linewidth}\leavevmode\epsfbox{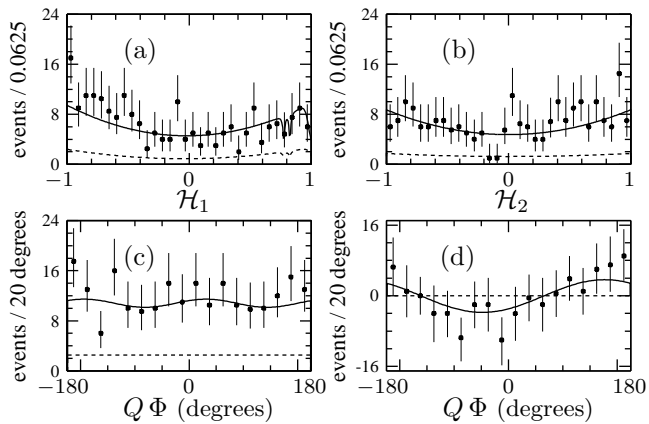}
}
\vspace{-0.3cm}
\caption{\label{fig:projection2} 
Projections onto the variables 
${\cal H}_1$ (a), ${\cal H}_2$ (b), $Q~\!\Phi$ (c), and
the differences between
the $Q~\!\Phi$ projections for events with 
${\cal H}_1~\!{\cal H}_2>0$ and with
${\cal H}_1~\!{\cal H}_2<0$ (d) 
for the signal $B^0\to\phi K^{*0}(892)$ candidates. 
}
\end{figure}
\begin{figure}[h]
\centerline{
\setlength{\epsfxsize}{1.0\linewidth}\leavevmode\epsfbox{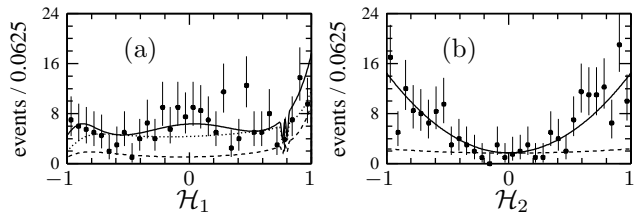}
}
\vspace{-0.3cm}
\caption{\label{fig:projection3} 
Projections onto the variables 
${\cal H}_1$ (a) and ${\cal H}_2$ (b)
for the signal $B^0\to\phi K^{*0}(1430)$ candidates. 
The difference between the solid and dotted lines in (a) shows 
the contribution of the tensor state to the angular distribution.
}
\end{figure}

We allow for multiple candidates in a given event by assigning 
to each a weight of $1/N_i$, where $N_i$ is the number of 
candidates in the same event. The average number of candidates 
per event is 1.04. The extended likelihood for a sample 
of $N_{\rm cand}$ candidates is 
\begin{equation}
{\cal L} = \exp\left(-\sum_{j}^{} n_{j}\right)\, 
\prod_{i=1}^{N_{\rm cand}} 
\exp\left(\frac{\ln{\cal L}_i}{N_i}\right).
\label{eq:likel}
\end{equation}
The event yields $n_j$, asymmetries ${\cal A}_j$,
and the signal polarization parameters $\vec{\alpha}$
are obtained by maximizing ${\cal L}$.


The results of our maximum likelihood fit to the sample
of $B^0\to\phi K^{*0}(892)$ candidates are summarized 
in Table~\ref{tab:results}. We also repeat the fit with
the requirement $1.13 < m_{K\!\pi} < 1.73$ (GeV) and without the
angular information. We observe $181\pm17$ events 
(statistical errors only) of the decays 
$B^0\to\phi K^{*0}(1430)$
with statistical significance greater than 10$\sigma$. 
In Fig.~\ref{fig:projection1}--\ref{fig:projection3}
we show projections onto the variables, where
data distributions are shown
with a requirement on the signal-to-background
probability ratio ${\cal P}_{\rm sig}/{\cal P}_{\rm bkg}$ 
calculated with the plotted variable excluded.
The solid (dashed) lines show the signal-plus-background
(background) PDF projections.

\begin{figure}[hbt]
\centerline{
\setlength{\epsfxsize}{1.0\linewidth}\leavevmode\epsfbox{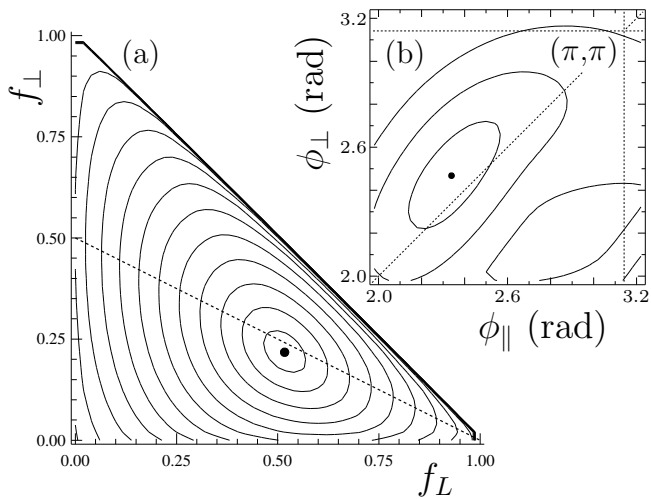}
}
\caption{\label{fig:contour} 
Likelihood function contours with 1$\sigma$ intervals 
for polarization (a) and phase (b) measurements
in the $B^0\to\phi K^{*0}(892)$ analysis. 
The fit results are shown with dots.
Diagonal dashed lines $f_\perp=(1-f_L)/2$ and 
$\phi_\perp=\phi_\parallel$ correspond to $|A_{+1}|\gg|A_{-1}|$.
In (b) the $(\pi, \pi)$ point is indicated by the crossed dashed lines.
}
\end{figure}

In the analysis of the decay $B^0\to\phi K^{*0}(892)$ 
for any given set of values ($\phi_{\parallel},\phi_{\perp},
\Delta\phi_{\parallel},\Delta\phi_{\perp}$) 
simple transformations of the angles, for example 
($-\phi_{\parallel},\pi-\phi_{\perp},
-\Delta\phi_{\parallel},-\Delta\phi_{\perp}$),
give rise to other sets of values which satisfy
Eq.~(\ref{eq:helicityfull}) in an identical manner.
To resolve this ambiguity, the set of values lying closest to 
the theoretical expectation 
($\pi,\pi,0,0$)~\cite{bvv1, bvv2, helicity} is chosen.
In Fig.~\ref{fig:contour}
we show likelihood function contour plots.

We find the decay $B^0\to\phi K^{*0}(1430)$
to be predominantly longitudinally polarized based on the 
${\cal H}_2$ angular distribution in
Fig.~\ref{fig:projection3} (b).
The width~\cite{pdg} and the angular distribution of the 
$K^{*0}(1430)$ resonance structure are not consistent with 
the pure $K_2^{*0}(1430)$ tensor state at more than 10$\sigma$. 
However, the angular distribution provides evidence 
(with statistical significance of 3.2$\sigma$) of the 
longitudinally polarized tensor $K_2^{*0}(1430)$
contribution in addition to the scalar $K_0^{*0}(1430)$,
see Fig.~\ref{fig:projection3} (a).


Our $B^0\to\phi K^{*0}(892)$ fit is performed with the 
$B\to f_0K^*$ and $B\to\phi{K}\pi$ contributions unconstrained. 
We obtain the event yields $25\pm10$ and $11\pm15$, respectively. 
The systematic uncertainties due to interference are
estimated using generated samples with conservative assumptions 
about the S-wave intensity and the interference phase. 
Additional systematic uncertainty originating from $B$ background
is taken as the difference between the fit results with the
combinatoric $\BB$ background component fixed to zero and fixed
to the expectation from~MC.

We vary the PDF parameters within 
their respective uncertainties, and derive the associated 
systematic errors.
The biases from the finite resolution of the helicity angle 
measurement and the dilution due to the presence of fake 
combinations are estimated with MC simulation.

The systematic errors in efficiencies are dominated 
by those in track finding and particle identification.
Other systematic effects arise from event-selection criteria, 
$\phi$ and $K^{*0}$ branching fractions, 
MC statistics, and number of $B$ mesons.
We calculate the efficiencies using the measured polarization 
and assign a systematic error corresponding to the total 
polarization uncertainty.
We find the uncertainty in the charge asymmetry due to the 
track reconstruction and identification to be less than~0.02.

In summary,
we have performed a full angular analysis and searched 
for $C\!P$ violation in the angular distribution with 
the decays $\BorBbar^0\to\phi\KorKbar^{*0}(892)$. 
Our results are summarized in Table~\ref{tab:results}. 
We observe, with more than 5$\sigma$ significance, 
non-zero contributions from all of the three amplitudes
$|A_0|$, $|A_\perp|$, and $|A_\parallel|$, see 
Fig.~\ref{fig:contour} (a). We also find 3$\sigma$ 
evidence for non-zero final-state-interaction phases, 
see Fig.~\ref{fig:contour} (b).
These results supersede our earlier measurements in this 
channel~\cite{babar:phikst, babar:vv}.
We also observe the decay $B^0\to\phi K^{*0}(1430)$.

For $B$ decays to light charmless particles
we expect the hierarchy of decay amplitudes to be 
$|A_0|\gg|A_{+1}|\gg|A_{-1}|$ 
under the assumption of pure loop diagram 
contribution, which is analogous to the discussion 
in Ref.~\cite{helicity}. 
Our measurements with the decay $B^0\to\phi K^{*0}(892)$
do not agree with the first inequality
but agree with the previous measurements in 
Ref.~\cite{babar:vv, belle:phikst}.
This suggests other contributions to the decay amplitude,
previously neglected, either within or beyond the 
Standard Model~\cite{bvv1, vvnew}.


We would like to thank Yuval Grossman, Alexander Kagan,
David London, Mahiko Suzuki, and Arkady Vainshtein 
for useful discussions.
We are grateful for the excellent luminosity and machine conditions
provided by our \pep2\ colleagues, 
and for the substantial dedicated effort from
the computing organizations that support \babar.
The collaborating institutions wish to thank 
SLAC for its support and kind hospitality. 
This work is supported by
DOE
and NSF (USA),
NSERC (Canada),
IHEP (China),
CEA and
CNRS-IN2P3
(France),
BMBF and DFG
(Germany),
INFN (Italy),
FOM (The Netherlands),
NFR (Norway),
MIST (Russia), and
PPARC (United Kingdom). 
Individuals have received support from CONACyT (Mexico), A.~P.~Sloan Foundation, 
Research Corporation,
and Alexander von Humboldt Foundation.


\bibliographystyle{h-physrev2-original}   

\begin{thebibliography}{99}

\bibitem{bvv1}
G.~Valencia, Phys.\ Rev.\ D {\bf 39}, 3339 (1989);
A.~Datta and D.~London, Int.\ J.\ Mod.\ Phys.\ A {\bf 19}, 2505 (2004).

\bibitem{cleo:phikst}
CLEO Collaboration, R.A.~Briere {\it et al.}, 
Phys.\ Rev.\ Lett. {\bf 86}, 3718 (2001).

\bibitem{babar:phikst}
$\babar$ Collaboration, B.~Aubert {\it et al.},
Phys.\ Rev.\ Lett. {\bf 87}, 151801 (2001);
Phys.\ Rev.\ D {\bf 65}, 051101 (2002).

\bibitem{babar:vv}
$\babar$ Collaboration, B.~Aubert {\it et al.},
Phys.\ Rev.\ Lett. {\bf 91}, 171802 (2003); arXiv:hep-ex/0303020.

\bibitem{belle:phikst}
BELLE Collaboration, K.-F. Chen {\it et al.},
Phys.\ Rev.\ Lett. {\bf 91}, 201801 (2003).

\bibitem{bvv2}
G. Kramer and W.F. Palmer, Phys.\ Rev.\ D {\bf 45}, 193 (1992);
H.-Y.~Cheng and K.-C.~Yang, Phys.\ Lett.\ B {\bf 511}, 40 (2001);
C.-H. Chen, Y.-Y. Keum, and H-n. Li, Phys.\ Rev.\ D {\bf 66}, 054013 (2002).

\bibitem{babar}
\babar\ Collaboration, B.~Aubert {\it et al.},
Nucl.\ Instrum.\ Methods {\bf A479}, 1 (2002).

\bibitem{pep} 
PEP-II Conceptual Design Report, SLAC-R-418 (1993).

\bibitem{bigPRD}
$\babar$ Collaboration, B.~Aubert {\it et al.},
to appear in Phys.\ Rev.\ D (2004), arXiv:hep-ex/0403025,
SLAC-PUB-10381.

\bibitem{geant} The $\babar$ detector Monte Carlo 
simulation is based on GEANT4:
GEANT4 Collaboration, S.~Agostinelli {\it et al.},
{Nucl.\ Instr.\ Meth.\xspace} A {\bf 506}, 250 (2003).

\bibitem{helicity} A.~Ali {\it et al.},
Z.\ Phys.\ C {\bf 1}, 269 (1979); 
M.~Suzuki, Phys.\ Rev.\ D {\bf 66}, 054018 (2002).

\bibitem{pdg}
S. Eidelman {\it et al.}, Phys.\ Lett.\ B {\bf 592}, 1 (2004).

\bibitem{vvnew} 
Y.~Grossman, Int.\ J.\ Mod.\ Phys.\ A {\bf 19}, 907 (2004),
A.~Kagan, arXiv:hep-ph/0405134;
P.~Colangelo, F.~De~Fazio, and T.N.~Pham, arXiv:hep-ph/0406162.

\end{thebibliography}

\end{document}